\documentclass{article}
\usepackage{spconf,amsmath,graphicx}
\usepackage{cite}
\usepackage{graphicx}
\usepackage{booktabs}
\usepackage{multirow}
\usepackage{amssymb}
\usepackage{color}
\usepackage{url}
\usepackage{comment}

\title{DDSP Guitar Amp: Interpretable Guitar Amplifier Modeling}
\name{\begin{tabular}{c}Yen-Tung Yeh$^{1, 2}$, Yu-Hua Chen$^{1, 2}$,  Yuan-Chiao Cheng$^{2}$, Jui-Te Wu$^{2}$, \\ 
Jun-Jie Fu$^{2}$, Yi-Fan Yeh$^{2}$, and Yi-Hsuan Yang$^{1}$\end{tabular}}
\address{$^{1}$ National Taiwan University, Taiwan, $^{2}$ Positive Grid, Taiwan}

\begin{document}
%
\maketitle

\begin{abstract}

Neural network models for guitar amplifier emulation, while being effective, often demand high computational cost and lack interpretability. Drawing ideas from physical amplifier design, this paper aims to address these issues with a new differentiable digital signal processing (DDSP)-based model, called ``DDSP guitar amp,'' that models the four components of a guitar amp (i.e., preamp, tone stack, power amp, and output transformer) using specific DSP-inspired designs.
With a set of time- and frequency-domain metrics, we demonstrate that DDSP guitar amp achieves performance comparable with that of black-box baselines while requiring less than 10\% of the computational operations per audio sample, 
thereby holding greater potential for usages in real-time  applications.

\end{abstract}
\section{Introduction}
\label{sec:intro}

A physical guitar amplifier often comprises four main components \cite{pakarinen2009review}: a) the \emph{preamp}, which boosts the  input signal and shapes its sound characteristics, often introducing deliberate distortion; b) the \emph{tone stack}, allowing frequency-specific adjustments; c) the \emph{power amp}, which nonlinearly amplifies the signal to drive speakers and can add further coloration; and d) the \emph{output transformer}, matching impedance and contributing additional nonlinearity. Each component introduces unique sound traits, with user-adjustable control knobs (such as ``gain,'' ``bass,'' ``mids,'' ``treble,'' and ``master'' volume) influencing the overall sound. 
There are intricate interactions between these components and their control knobs.

Guitar amplifier modeling 
aims to digitally replicate the 
behavior of physical amps
and create similarly wide range of tone variations.
The task is a type of virtual analog (VA) modeling, which aims to digitally replicate real-world devices. In general, approaches to VA modeling  span a spectrum from white-box to black-box methods, based on their reliance on circuit knowledge. White-box approaches \cite{pakarinen2009review, diffWhiteBox, Parker:2022.PhysicalModeling, 5280324} require comprehensive circuit understanding; grey-box methods \cite{eichas2018jaes, miklanek2023neural, nercessian2021lightweight, eichas2017block, HMModel} utilize partial device knowledge, balancing interpretability and flexibility; and black-box approaches \cite{app10020638, Damskgg2019RealTimeMO,8683529,steinmetz2022efficient, 8682805,6567472,schmitz2018real,Zhang2018AVG,769f627fa4fe49569bd207f6b1d32dc3, 10094769, chen24dafx, chen2024zeroshotamplifiermodelingonetomany,yeh24dafx}, in particular those based on deep neural networks (NN), have recently demonstrated impressive abilities in capturing complex nonlinear behaviors of amps and beyond using only input-output data, without prior circuit knowledge.

Despite its prevalence, existing black-box NN-based approaches have notable downsides.
For example, while large NN models tend to perform well \cite{Damskgg2019RealTimeMO, 769f627fa4fe49569bd207f6b1d32dc3}, the incurred computational cost limits their portability to low-power devices and they may suffer from system instability in real-time usage \cite{vanhatalo2022review}. The lack of interpretability is also an important issue, making it hard for musicians to intuitively fine-tune parameters for specific sonic characteristics \cite{vanhatalo2022review}. 

\begin{figure}[tb]
\begin{minipage}[b]{1.0\linewidth}
  \centering
  \centerline{\includegraphics[width=7cm]{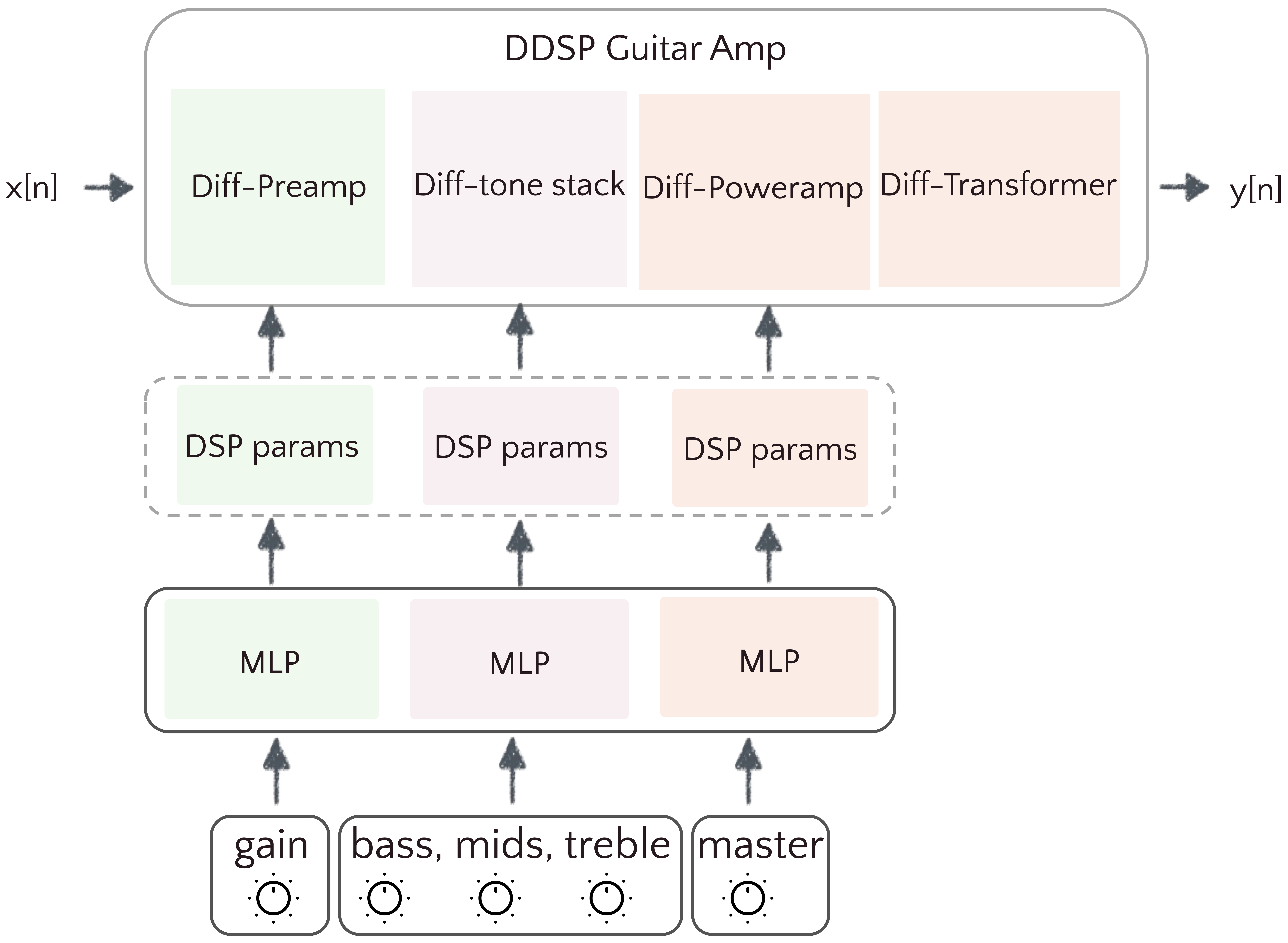}}
\end{minipage}
\caption{Illustration of the proposed DDSP-based model for guitar amplifier modeling, named DDSP guitar amp.}
\label{fig:system}
\end{figure}

Drawing inspirations from existing grey-box VA modeling methods and the general idea of differentiable digital signal processing (DDSP) \cite{engel2020ddsp,renault2023ddsp_piano}, this paper aims to propose a novel grey-box DDSP-based approach that offers a solution for efficient and interpretable guitar amp modeling.
Specifically, we follow closely the physical design of guitar amps and propose differentiable modules corresponding to the four main physical components of an amp.
As depicted in Figure \ref{fig:system}, the four components are cascaded in the same way as their physical counterparts, and are conditioned by parameters derived from human-readable control knobs through a few dense multilayer perceptron (MLP) layers.
To our best knowledge, this represents the first attempt to neural guitar amp modeling that uses such a modularized and DSP-inspired design.
We refer to our model as ``DDSP guitar amp.'' 

Two works are highly related to ours.
Eichas \emph{et al.} \cite{eichas2018jaes} 
presented a grey-box method for modeling guitar amps 
with the Wiener-Hammerstein (WH) model rooted in nonlinear system identification \cite{7152767,WHMODEL}, achieving better interpretablity. 
However, this work is preliminary in that they only modeled the preamp, 
ignoring the other components of an amp. Moreover, they did not consider control knobs. 
Mikl\'anek \emph{et al.} \cite{miklanek2023neural} employed circuit methods to model instead the tone stack component of guitar amps, with a mechanism to account for the control knobs related to the tone stack. However, for the other components (e.g., preamp and power amp), they simply used a black-box NN to model them. 
A comprehensive approach that remains interpretable for all the four components of a guitar amp and that 
captures the full range of knob-induced variations as presented here has yet to be developed.

Our approach has several key innovations: 1) it captures intricate knob-output relations across all amp components via end-to-end training, overcoming the limited parameter configs of previous interpretable models; 2) it maintains interpretability through the DDSP principles, enabling intuitive parameter adjustments that correspond to physical amp characteristics; and 3) it achieves performance comparable to 
or surpassing 
current black-box NN models, with less than 10\% of the operations per audio sample, 
making it 
suitable for real-time applications
in music production and performance.

We provide audio examples at \url{https://ytsrt66589.github.io/ddspGuitarAmp_Demo/}.  

\section{Proposed methods}
\label{sec:methods}

Our objective is to develop a model $g$ that accurately emulates a target guitar amplifier. For an input signal $x[n]$ and control parameters $\phi$ representing knob settings, the model produces an output $y[n] = g(x[n], \phi)$ that closely matches the acoustic characteristics of the output of the physical device. 
We provide model details below.

\subsection{Differentiable Preamp}
\label{sec:methods:preamp}

The preamp, the first stage in a guitar amplifier, boosts the input signal and determines its initial tone \cite{pakarinen2009review}. It exhibits dynamic nonlinear behavior significantly influenced by its ``bias point'' \cite{pakarinen2009review, eichas2018jaes}, 
which is determined by the DC operating conditions of the circuit. 
The bias point would vary with the input signal level,
leading to dynamic variations in harmonic content and compression characteristics. The ``gain" knob affects the preamp behavior by altering both the signal amplification and frequency response, resulting in diverse sound variation.

Inspired by \cite{eichas2018jaes},
we adapt the Wiener-Hammerstein (WH) model \cite{WHMODEL,7152767} to capture this complex behavior. Here, a ``WH block'' is composed of two linear filters $H_1$ and $H_2$, each implemented as cascades of low-shelf, peak, and high-shelf filters, and a nonlinear function $f(\cdot)$:
\begin{equation}
\label{equation:WH}
z'[n] = H_2(f(H_1(z[n])))\,,
\end{equation}
where $z[n]$ and $z'[n]$ denote the input and output of a block.

Different from \cite{eichas2018jaes}, which uses a static nonlinear function such as  $\tanh$ for $f(\cdot)$,
we propose to use a gated recurrent unit (GRU) \emph{with hidden size 1} as $f(\cdot)$ to specifically account for the behavior of the bias point in the preamp. This innovative use of a GRU effectively models the signal level-dependent nonlinearity and captures the dynamic response of the preamp. By leveraging the ability of GRU to learn complex nonlinearities from data, we maintain interpretability by emulating the ``memory'' effect in tube circuits \cite{pakarinen2009review}. Crucially, we find additionally that our GRU-based approach accurately represents \emph{asymmetric distortion}, a key characteristic of tube preamps that cannot be modeled by symmetric functions such as $\tanh$. 
While there are methods to introduce asymmetry to $\tanh$ functions (e.g., adding DC offsets or combining multiple $\tanh$), these often require careful tuning and may not capture the full complexity of the preamp's behavior. Moreover, unlike GRU, $\tanh$ is not able to model the time-dependent behaviors of preamps, since the output of  $\tanh$ is not affected by the previous output. 
Finally, we note that, despite harnessing neural network capabilities, the GRU maintains computational efficiency suitable for real-time applications.

To  enhance the model's versatility, we employ $N=4$ cascaded WH blocks. 
Moreover, as shown in Figure \ref{fig:system_details}a,
we incorporate information from the ``gain'' knob both before and after the nonlinear stage $f(\cdot)$ in each WH blck. 
The ``pregain,'' applied before the nonlinear stage, controls the amount of signal fed into the nonlinearity, effectively determining the level of distortion. The ``postgain,'' applied after the nonlinear stage, adjusts the overall output level without altering the distortion characteristics. This  allows for fine-tuned control over the distortion amount and overall volume independently, enabling accurate reproduction of various distortion levels and acoustic characteristics across different gain settings. 

\subsection{Differentiable Tone Stack}
\label{sec:methods:tonestack}

The tone stack, positioned after the preamp, is crucial for shaping the amplifier's frequency response. It typically features three interactive controls: ``bass,'' ``mids,'' and ``treble,'' allowing guitarists to tune and sculpt the preamp's distorted output \cite{yeh2008simulating}. The complex interdependence of these controls creates a nuanced tone shaping system.

The pursuit of a differentiable tone stack has been done before by Mikl\'anek \emph{et al.} \cite{miklanek2023neural}. 
Similarly,
we model it as a series of three filters: low-shelf, peak, and high-shelf (LPH)
\cite{reiss2014audio}, as shown in Figure \ref{fig:system_details}b. This structure effectively replicates various tone stack configs by adjusting filter parameters based on knob settings. 
The low-shelf and high-shelf filters primarily modulate bass and treble responses, respectively, while the peak filter targets mid-range frequencies. 
Our digital representation not only mimics the frequency response of analog tone stacks but also preserves the critical interactive relationships between the control knobs,
capturing the 
subtleties
that characterize different guitar amplifiers.

\subsection{Differentiable Power Amp}
\label{sec:methods:poweramp}

The power amp is the last amplification stage before the output transformer.
Its behavior is affected by output tube type, class of operation, and negative feedback \cite{708439}. 
Little work, however, has been done to build differential power amp and output transformer. We present such an endeavor below.

We pioneer the design of a differentiable power amp following the push/pull topology (POW) in the physical design of a power amp, 
which 
includes a phase splitter, a negative feedback loop, and a presence control \cite{pakarinen2009review}. Our DSP 
implementation of the phase splitter (\texttt{PS}) inverts one audio segment with phase inversion (\texttt{PI}) and applies a simplified soft-clipper nonlinear function, approximating the behavior of tube phase splitters while maintaining computational efficiency \cite{dempwolf2009influence}. Implementing negative feedback in digital systems traditionally introduces echo artifacts due to the inherent delay. We address this challenge by using a linear filter to emulate negative feedback behavior, effectively reducing distortion and controlling frequency response before the phase splitter stage 
with light computational cost. 
As shown in Figure \ref{fig:system_details}c, 
the signal path before \texttt{PS} 
includes a ``master'' volume control (\texttt{M}), a filter (\texttt{F}) emulating negative feedback and presence control, along with a gain control (\texttt{G}). After \texttt{PS}, both separate paths for each phase employ WH blocks with gain control and similarly a GRU with hidden size 1 for the nonlinear function.

\subsection{Differentiable Output Transformer}
\label{sec:methods:transformer}

The output transformer, 
which should not be confused with self-attention based transformers used in NNs \cite{vaswani2017attention}, performs impedance matching between the power amp and speaker, nonlinearly amplifying the signal and adding tone coloration. This behavior is primarily due to magnetic domain hysteresis and frequency-dependent losses \cite{bertram1994theory, cauduro2011real}. 
Moreover,
the output transformer exhibits a bandpass characteristic \cite{cauduro2011real}. 

\begin{figure}[tb]
\begin{minipage}[b]{1.0\linewidth}
  \centering  \centerline{\includegraphics[width=8.5cm]{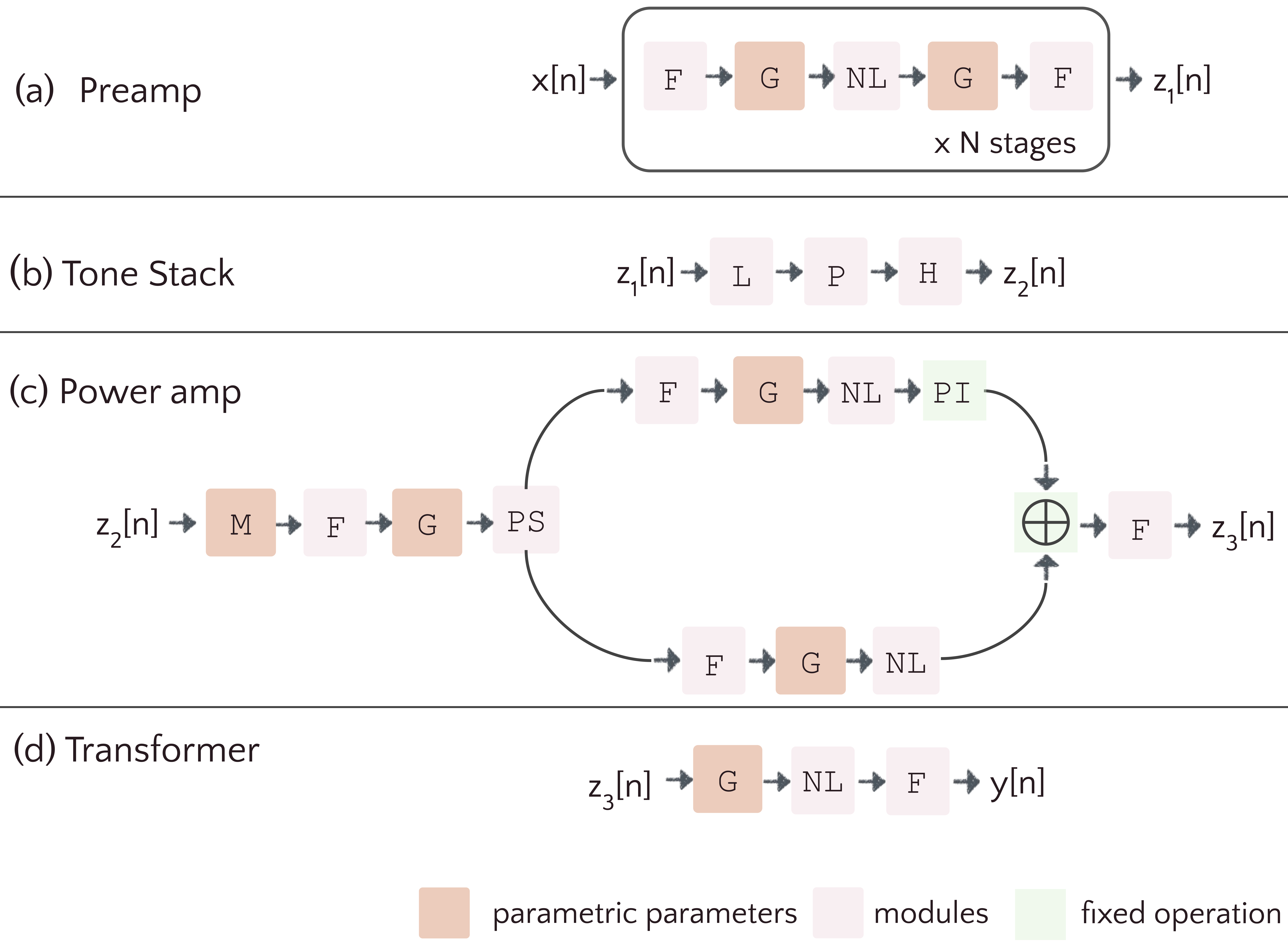}}
\end{minipage}
\caption{Details of DDSP Guitar Amp. Letters in the squares 
denote their respective types---\texttt{F}: filter, \texttt{G}: gain, \texttt{NL}: nonlinear function, \texttt{L}: low-shelf filter, \texttt{P}: peak filter, \texttt{H}: high-shelf filter, \texttt{M}: master, \texttt{PS}: phase splitter, \texttt{PI}: phase inversion. Orange blocks: knob controller-estimated parameters for multiplication. Pink blocks: designed operations with Knob Controller-predicted parameters. Green blocks: fixed operations without learnable parameters. Best viewed in color.}
\label{fig:system_details}
\end{figure}

We propose to model the output transformer (TRANS) simply with a GRU with again hidden size 1, complemented by gain control and filtering,
as shown in Figure \ref{fig:system_details}d. The GRU's state-dependent output captures the history-dependent nature of hysteresis, while the additional modules model the 
frequency-shaping effects. This approach, chosen over traditional models like Jiles-Atherton \cite{holters2016circuit}, offers data-driven parameter learning ability with light computational cost,
capturing the dynamic, nonlinear, and frequency-dependent characteristics of real transformers.
Interestingly, as demonstrated on our project page, we find that the resulting distortion curve closely resembles a standard magnetic hysteresis curve.

\subsection{Knob Controller}

To achieve full dynamic modeling, we implement MLPs as knob controllers. Each amp component has an independent set of MLP layers that map knob values to DSP parameters, enabling the model to learn complex, nonlinear relations between knob adjustments and sonic outcomes. 
This approach captures subtle interactions beyond simple scaling. For instance, the ``gain'' knob effect combines filtering and amplification, while the ``bass'' knob influences multiple frequency ranges. Learning these intricate relationships from data, our model has the opportunity to replicate the full spectrum of tone variations possible in physical guitar amps.

\begin{table*}[ht]
  \centering
  \begin{tabular}{lccccrr}
  \toprule
  \multirow{2}{*}{Model} & \multicolumn{2}{c}{Seen knob conditions} & \multicolumn{2}{c}{Unseen knob conditions}  & \multirow{2}{*}{Ops/sample} &\multirow{2}{*}{Params} \\
  \cmidrule(lr){2-5} 
  & MAE~$\downarrow$ & MR-STFT~$\downarrow$ & MAE~$\downarrow$ & MR-STFT~$\downarrow$ \\
  \midrule
   \emph{A}. Small Concat-GRU-8 & 0.057 & 4.302 & 0.075 & 5.762 & 1,344 & 369\\
   \emph{B}. Big Concat-GRU-48 & 0.013 & 1.214 & 0.023 & 1.851 & 19,872 & 7,969\\
  \midrule
  \emph{C}. WH Only & 0.317 & 2.552 & 0.189 & 4.675 & 736 & 4,462 \\
   \emph{D}. WH$+$LPH$+$WH & 0.063 & 5.098 & 0.066 & 5.803 & 995 & 10,213 \\
   \emph{E}.  WH$+$LPH$+$POW & 0.034 & 2.979 & 0.057 & 4.825 & 1,243 & 8,200 \\
   \emph{F}.  WH$+$LPH$+$POW$+$TRANS & 0.024 & 2.161 & 0.043 & 3.972 & 1,352 & 10,126 \\
  \bottomrule
  \end{tabular}
  \caption{Evaluation results of (\emph{A}--\emph{B}) black-box baselines and (\emph{F}) the proposed DDSP model and (\emph{C}--\emph{E}) its ablations.}
\label{tab:model_results}
\end{table*}

\section{Experimental Setup}
\label{sec:experiments}


Our experiment targets the emulation of Marshall JVM 410H, a tube-based guitar amplifier. We use the data from Miklanek \emph{et al.} \cite{miklanek2023neural}, comprising 6-minute audio files of unprocessed dry signals sampled at 44.1kHz, featuring diverse guitar and bass playing styles. Target signals were recorded from the ``OD1'' channel, a high-distortion setting.
We split the data into train, validation and test sets with a 6:1:3 ratio.
Following \cite{miklanek2023neural}, we use both seen and unseen knob conditions for testing. 


For performance comparison, we establish two black-box NN baselines: Small Concat-GRU-8 and Big Concat-GRU-48, with hidden sizes of 8 and 48 respectively, using the concatenation method \cite{yeh24dafx} for knob value conditioning. Moreover, 
we evaluate ablated versions of the proposed DDSP-based model, listed below with increasing complexity:
\begin{itemize}
  \item \textbf{WH only}: A single multi-block Wiener-Hammerstein model as described in Section \ref{sec:methods:preamp} that emulates the entire guitar amp, which resembles the approach of \cite{eichas2018jaes}.
  
  \item \textbf{WH+LPH+WH}: A WH model for the preamp, followed by a series of low-shelf, peak, high-shelf filters (LPH)  emulating the tone stack, and another WH model that emulates both the power amp and transformer altogether.
  This resembles the approach of \cite{miklanek2023neural} as the same NN structure are used before and after the tone stack.
  
  \item \textbf{WH+LPH+POW}: Replace the WH after LPH by the proposed 
  POW-based design described in Section \ref{sec:methods:poweramp}, omitting the transformer modeling in Section \ref{sec:methods:transformer}.

  \item \textbf{WH+LPH+POW+TRANS}: Our full proposed model.
\end{itemize}


Here are some implementation details.
We implemented all the models using the  \texttt{PyNeuralFx} library \cite{yeh2024pyneuralfxpythonpackageneural} and trained them on a single NVIDIA RTX 4090 GPU. Our training process  starts with an initial learning rate of 2e--3 over 100 epochs, adaptively halving the learning rate after two consecutive non-improving validations and early stopping after four.
The DDSP models were trained on 8,192-sample audio segments, while the GRU baselines used 2,048-sample segments due to gradient updating constraints in recurrent networks. We used mean absolute error (MAE) and multi-resolution STFT (MR-STFT) losses \cite{steinmetz2022efficient}. The STFT loss employs three FFT window sizes: 128, 512, and 2,048. For DDSP models, MLP controller outputs were sigmoid-activated and mapped to predefined parameter ranges, ensuring physically meaningful bounds.
All MLPs in our model comprise three layers with LeakyReLU activations (slope 0.1) and 32 hidden units. Condition values were normalized to $[-1, 1]$ for consistent input scaling. Linear filters in DDSP models were implemented as IIR biquad filters using \texttt{dasp-pytorch} \cite{dasp}. 
We employed the frequency sampling method for efficient biquad IIR filter updating as in \cite{nercessian2020neural, colonel2022direct}.
This balances model performance with computational efficiency, enabling fair comparison between our DDSP-based method and the baselines.

\section{Objective Evaluation Results}
\label{sec:results}


We evaluate the accuracy of the implemented models for emulating the target guitar amp (for unseen content in test data) in MAE  and MR-STFT \cite{yeh24dafx}.
Moreover, we assess model efficiency by operations (ops) per sample. For reference, sigmoid and tanh operations cost 30 floating point ops \cite{parker2019modelling}.


Table \ref{tab:model_results} presents the result, starting with the two baseline GRU models: (\emph{A}) Small Concat-GRU-8 and (\emph{B}) Big Concat-GRU-48. Config  \emph{B}, which is a strong baseline, attains the lowest MAE and MR-STFT for both seen and unseen conditions, but with significantly higher computational cost. Config  \emph{A} demonstrates that merely reducing GRU size is ineffective for balancing modeling accuracy and efficiency.

We then progressively evaluate four configs  (\emph{C}--\emph{F}) to validate our design choices. Config \emph{C} (WH model only) excels in frequency content reconstruction but struggles with time-domain aspects and lacks interpretability due to entangled components. Config \emph{D} introduces a structured approach separating preamp, tonestack, and power amp, but its simple WH structure for the power amp yields suboptimal results. Config \emph{E} improves upon config \emph{D} with a more sophisticated power amp model, showing noticeable enhancements. Finally, the proposed method (config \emph{F}), which incorporates an output transformer model, crucially enables performance comparable to config \emph{A} with improved accuracy and interpretability. Notably, config \emph{F} requires less than 10\% of the operations per sample compared to config \emph{B}.

The clear reduction in MAE and MR-STFT going from  \emph{C} to \emph{F} for either seen or unseen conditions validates our design process. While config \emph{B} has the lowest losses, config \emph{F} exhibits a better balance of accuracy, computational efficiency, and interpretability. Config \emph{F} outperforms the similarly-sized config \emph{A}, 
indicating the superiority of our physically-inspired approach over the simple approach of model scaling down.

\section{Conclusion}
\label{sec:conclusion}

In this paper, we have presented a DDSP-based method for guitar amplifier modeling, leveraging physical design principles to balance interpretability, computational efficiency, and accurate reproduction of nonlinear behavior. Our approach demonstrates advantages over black-box NN baselines in capturing component interactions and user control. Future work could explore addressing aliasing in nonlinear digital systems. 


\vfill\pagebreak

{\fontsize{9}{9.5}\selectfont
\bibliographystyle{IEEEbib}
\bibliography{refs}
}
\end{document}